\documentclass[journal]{IEEEtran}
\usepackage{amsmath, subfigure}
\usepackage{graphicx}
\usepackage{amssymb}
\usepackage{color}
\usepackage{longtable}
\usepackage{float}
\usepackage{epstopdf}
\usepackage{multirow}
\usepackage{bm}
\usepackage{cite}
\graphicspath{{figure/}}

\newtheorem{rmk}{\bf Remark}

\begin{document}

\title{Energy Imbalance Management Using a Robust Pricing Scheme}

\author{Wei-Yu~Chiu,~\IEEEmembership{Member,~IEEE}, Hongjian Sun,~\IEEEmembership{Member,~IEEE},  and
        H.~Vincent~Poor,~\IEEEmembership{Fellow,~IEEE}
\thanks{W.-Y. Chiu and H.~V.~Poor are with the Department of Electrical Engineering, Princeton University, Princeton, NJ 08544, USA (e-mail: wychiu@ieee.org; poor@princeton.edu).}%
\thanks{H. Sun is with the Division of Engineering, Kings College London, WC2R
2LS London, U.K. (e-mail: mrhjsun@hotmail.com).}
\thanks{This work was supported in part by the National Science Council of Taiwan under Grant NSC100-2917-I-564-014,
in part by the UK Engineering and Physical
Sciences Research Council (EPSRC) under Grant No. EP/I000054/1,
and
in part by the U.S. Air Force Office of Scientific Research under MURI Grant FA9550-09-1-0643.}
\thanks{
\copyright 2012 IEEE. Personal use of this material is permitted. Permission from IEEE must be obtained for all other uses, in any current or future media, including reprinting/republishing this material for advertising or promotional purposes, creating new collective works, for resale or redistribution to servers or lists, or reuse of any copyrighted component of this work in other works.
}
\thanks{Digital Object Identifier 10.1109/TSG.2012.2216554}
}
\maketitle

\begin{abstract}
This paper focuses on the problem of energy imbalance management in a microgrid. The problem is investigated
from the power market perspective. Unlike the traditional power grid, a microgrid can obtain extra energy from a renewable energy source (RES) such as a solar panel or a wind turbine. However, the stochastic input from the RES brings difficulty
in balancing the energy.
In this study, a novel pricing scheme is proposed that provides robustness against the intermittent power input.
The proposed scheme considers possible uncertainty in the marginal benefit and the marginal cost of the power market.
It uses all available information on the power supply, power demand, and imbalanced energy.
The  parameters of the scheme are evaluated using
an $H_{\infty}$ performance index.
It turns out that
the parameters
can be obtained by solving a linear matrix inequality problem, which is efficiently solvable due to its convexity.
Simulation examples are given to show its excellent performance in comparison with
existing area control error pricing schemes.
\end{abstract}


\section{Introduction}

Price is an important element of market behavior and is
closely related to energy consumption \cite{PC}, energy management \cite{P_E1}, load control \cite{LC}, etc.
A pricing scheme can be employed to balance the rate of change of the energy resources \cite{Alva0}.
In a power market, the power demand and supply are associated with the
market price: from the consumers' perspective, the demand increases/decreases as
the marginal benefit is higher/lower than the price; from the suppliers' perspective,
the supply increases/decreases when
the marginal cost is lower/higher than the price. For a functional pricing scheme,
changing the market price can control the energy imbalance.

Many studies have investigated the power market behavior from a system perspective, i.e.,
by examining the power market dynamics \cite{Alva1,Alva2,Market_clear,Com_re}.
In general, the power market dynamics at least consist of power demand dynamics and power supply dynamics.
When energy storage is considered, the power market model also included the power storage dynamics \cite{Alva1}.
To balance the energy, i.e., to drive the energy storage to zero, a pricing scheme termed area control error (ACE) pricing scheme was studied in \cite{Alva2,Market_clear,Com_re}.
The ACE pricing scheme uses  feedback about the energy imbalance  to control the rate of change of the price.
In the terminology of control theory, this ACE pricing scheme is a dynamic pricing controller.

In this paper, we pay particular attention to the power market for a microgrid, which is different from the scenarios considered
in the existing studies \cite{Alva1,Alva2,Market_clear}.
Microgrids, also termed distributed resource island systems,
are defined as ``all intentional island systems that could include local and/or area electric power systems'' \cite{micro_def}.
For the purposes of this study,
a microgrid can be any smart facility or unit that efficiently uses energy to maintain its smart functionality.
Meanwhile, it can acquire extra power input from a local renewable energy source (RES), e.g.,
 solar panels or wind turbines.
In this case, the overall power supply to the consumer is different from the case considered in  \cite{Alva1} and \cite{Alva2},
where the power input  comes only from power suppliers, e.g., power companies.

Although a microgrid can use the energy efficiently, one significant challenge encountered by employing RESs
is the intermittent (or stochastic, fluctuating) power input to the grid \cite{wind}. This intermittent attribute results from unpredictable
weather conditions. From a perspective of energy management, it causes difficulty in balancing the power demand and
the power supply. Traditionally,
the ACE pricing scheme \cite{Alva1,Alva2} controls the rate of change of the price  so that
the rate is proportional to the negative value of the imbalanced energy.
By doing so, the imbalanced energy can be well managed.
In this study, we reveal that its performance can degrade when  an extra intermittent power input is involved.
Therefore, a  pricing scheme that is robust against fluctuating power input is needed.

This paper extends the power market model studied in \cite{Alva1,Alva2,Market_clear,Com_re} to a generalized scenario by including the uncertainty in the marginal benefit and the marginal cost.
This extension results in a stochastic power system.
We propose a novel pricing scheme for energy imbalance management by using
 fuzzy interpolation techniques \cite{Hinf}.
 To combat the uncertainty and the fluctuating power effects from the RES, an $H_{\infty}$ performance index
is adopted \cite{Robust_C,Hinf}: the proposed pricing scheme is designed such that the imbalanced energy over all possible
disturbances, i.e., the uncertainty and the fluctuating effects, is less than a fixed attenuation level.
  The pricing parameters can then be obtained by solving a linear matrix inequality (LMI) \cite{Boyd_LMI}, which is convex and thus is efficiently solvable \cite{Boyd_conv}.

The main contributions of this paper are as follows. This study proposes a pricing design
from a system perspective that allows further extension to a more complicated
power market system. In contrast to existing pricing schemes \cite{Alva1,Alva2},
the proposed scheme is more general and robust as it
considers system disturbances, especially the uncertain and fluctuating effects of RESs.
Based on the proposed methodology,
it is found that the price vibration plays an important role in balancing the energy excess or energy deficiency.
Simulations show that the proposed scheme outperforms existing ACE pricing schemes
both in the traditional setting and the scenario studied in this paper.

To avoid confusion, this paper adopts standard notation. Lowercase letters, such as $p_g, p_d, e$ and $\gamma$, represent scalars;
bold and lowercase letters, such as $\bm{x},\bm{y}$ and $\bm{b}$, represent vectors; bold and capital letters, such as $\bm{A},\bm{K},\bm{C}$ and $\bm{D}$, are used to denote matrices. For convenience, $\bm{I}_n$ denotes the $n\times n$ identity matrix. $\bm{A}^T$ denotes the transpose of  $\bm{A}$.
For a function $f(t)$ depending on time $t$, $\dot{f}(t)$ denotes the derivative of $f(t)$ with respect to $t$, i.e., $\dot{f}(t)=\frac{df(t)}{dt}$.
The notation $\bm{A}\succ0$ is used to denote
 a symmetric and positive-definite matrix $\bm{A}$, i.e.,
$\bm{A}^T=\bm{A}$   and $\bm{x}^T\bm{A}\bm{x}>0$ for all $\bm{x}\not=0$. Meanwhile,
$\bm{B}\prec 0$ means that $-\bm{B}\succ 0$.
For a symmetric matrix $\bm{A}$, ``$\star$'' is used to denote symmetric terms in $\bm{A}$, e.g.,
$[\bm{A}]_{ij}=\star$ implies $[\bm{A}]_{ij}=[\bm{A}]_{ji}$, where $[\bm{A}]_{ij}$ represents the $(i,j)$-entry of $\bm{A}$.

The rest of this paper is organized as follows. Section \ref{sec_ProbFor} formulates the power market dynamics and
extends the model in \cite{Alva1} and \cite{Alva2} to include  market system disturbances. The proposed pricing scheme is described in
Section \ref{sec_proposed}.  Simulation results are presented in Section \ref{sec_sim}. Finally, Section \ref{sec_con} concludes this paper.

\section{System Dynamics of Power Supply, Power Demand, and Energy Storage}\label{sec_ProbFor}

This section presents the system dynamics of the microgrid power market, including the dynamics of
power demand, power supply and energy storage.
A microgrid needs a certain amount of power to maintain its smart functionality.
The required power, denoted by  $p_d(t)$, comes from the connected RES, denoted by $in(t)$, and a power supplier, denoted by $p_g(t)$.
To balance the energy, it is desirable to have the power demand $p_d(t)$ equal to the sum of  $in(t)$ and $p_g(t)$.
In this structure, the power demand $p_d(t)$ relates to the current price  $\lambda(t)$ and its marginal benefit,
and the power supply depends on the power generation cost, the market price, and feedback information about the previous excess power. The goal is to design a pricing scheme $\lambda(t)$ that can balance the energy.
In other words, we want to stabilize the imbalanced energy $e(t)$, i.e., to drive the stored energy $e(t)$ to zero.

To study the power market dynamics, Alvarado's model \cite{Alva1,Alva2} is considered.
In this paper, this model is  extended to involve the fluctuating power input from the RES, and
the uncertainty in marginal cost and the marginal benefit.
We will briefly discuss Alvarado's  power market model, and the reader can refer to \cite{Alva1,Alva2,Market_clear}  for further details.
 For simplicity, the case of a single supplier and a single consumer, which was the focus of \cite{Market_clear},
 is considered.
Let $p_g(t)$ be the power supply (or power generation) to the microgrid at time $t$.
 The corresponding marginal cost for supplying  $p_g(t)$ is denoted by
 $b_g+c_g p_g(t)$, where $b_g$ and $c_g$ represent the initial supplier cost and the supplier's demand elasticity, respectively \cite{Alva1}. For an economic system, the power supply  speed $\dot{p}_g(t)$ increases as the price  $\lambda(t)$
 exceeds the cost $b_g+c_g p_g(t)$, while it decreases as the price $\lambda(t)$ is lower than  $b_g+c_g p_g(t)$.
 Based on Alvarado's model, the speed
$\dot{p}_g(t)$  can be expressed in terms of the marginal cost and the price as
\begin{equation}\label{eq_p_g}
   \dot{p}_g(t)=\frac{1}{\tau_g}\times \{ \lambda(t)- (b_g+c_g p_g(t))-ke(t) \}
\end{equation}
where $\tau_g$ is a scale factor and  $e(t)$ represents the stored energy.
The extra term $ke(t)$ with $k>0$ is considered as the additional cost for the excess power supply.
It is essential to include $ke(t)$ in (\ref{eq_p_g}) to ensure stability.

 As suggested by \cite{Alva1}, the dynamic model (\ref{eq_p_g}), which is referred to as the power supply dynamics in this paper, may involve some uncertain or stochastic attributes.
The uncertainty can be
 presented by the
term $b_g$. In this case,
we regarded $b_g$ as a random process by considering
\begin{equation}\label{eq_bg}
b_g=\hat{b}_g+\Delta_g(t)
\end{equation}
 where $\hat{b}_g$ represents a known nominal value (average value) of  $b_g$, and $\Delta_g(t)$ models the uncertainty.
The power supply dynamics  can then be rewritten by using (\ref{eq_p_g}) and  (\ref{eq_bg}) as
\begin{equation}\label{eq_p_g2}
  \dot{p}_g(t)=  \frac{-c_g}{\tau_g}p_g(t)  -\frac{k}{\tau_g}e(t)  - \frac{\hat{b}_g}{\tau_g}  +\frac{1}{\tau_g}\lambda(t)-\frac{1}{\tau_g}\Delta_g(t).
\end{equation}

Let us consider the power demand of the microgrid.
The initial consumer benefit and the consumer's demand elasticity
are denoted by $b_d$ and $c_d$, respectively.
Analogously to the relation between the power supply and the market price,
 the demand rate $\dot{p}_d(t)$ increases if the marginal benefit  $b_d+c_d p_d(t)$ exceeds the price $\lambda(t)$, and
 the rate declines as  $\lambda(t)\geq b_d+c_d p_d(t)$. Thus the power demand dynamics can be described by \cite{Alva1}
\begin{equation}\label{eq_p_d0}
      \dot{p}_d(t)=\frac{1}{\tau_d}\times \{(b_d+c_d p_d(t))-\lambda(t) \}
\end{equation}
where $\tau_d$ is a scale factor.
To model the stochastic uncertainty as in (\ref{eq_bg}), $b_d$ is replaced by
 $\hat{b}_d+\Delta_d(t)$ and hence, the demand dynamics (\ref{eq_p_d0}) can be reformulated as
 \begin{equation}\label{eq_p_d}
  \dot{p}_d(t)=  \frac{c_d}{\tau_d}p_d(t) + \frac{\hat{b}_d}{\tau_d}  -\frac{1}{\tau_d}\lambda(t)+  \frac{1}{\tau_d} \Delta_d(t).
\end{equation}

As the RES can produce energy, an extra power input $in(t)$ is available to the microgrid.
In contrast to $p_g(t)$, which is a steady power source that relates to the marginal cost, $in(t)$ does not contribute to the cost but provides an intermittent power gain.
For the power supply $p_g(t)$, the power demand $p_d(t)$ and the power input $in(t)$, the power imbalance  $\dot{e}(t)$ (the derivative of the stored energy)
can be formulated as
 \begin{equation}\label{eq_e}
  \dot{e}(t)=  p_g(t) + in(t) - p_d(t).
\end{equation}
The goal is to find a pricing scheme $\lambda(t)$, affecting the dynamics in  (\ref{eq_p_g2}) and (\ref{eq_p_d}),  such that
the imbalanced energy $e(t)$ can be driven to zero.

For convenience, we define
\begin{equation*}\label{eq_def}
\begin{split}
& \bm{x}(t)= \; [p_g(t)\;p_d(t)\;e(t)]^T, \bm{b}=[- \frac{\hat{b}_g}{\tau_g}\;\frac{\hat{b}_d}{\tau_d}\;0 ]^T,\\
& \bm{w}(t)= \; [\Delta_g(t)\;\Delta_d(t)\;in(t)]^T , \bm{\tau}=[ \frac{1}{\tau_g}\; \frac{-1}{\tau_d}\; 0  ]^T,\\
& \bm{A}= \;
\left[
  \begin{array}{ccc}
    -\frac{c_g}{\tau_g} & 0 & -\frac{k}{\tau_g} \\
    0 & \frac{c_d}{\tau_d} & 0 \\
    1 & -1 & 0 \\
  \end{array}
\right],\mbox{ and }
 \bm{B}= \;
\left[
  \begin{array}{ccc}
    -\frac{1}{\tau_g} & 0 & 0 \\
    0 & \frac{1}{\tau_d} & 0 \\
    0 & 0 & 1 \\
  \end{array}
\right].
\end{split}
\end{equation*}
Based on (\ref{eq_p_g2}), (\ref{eq_p_d}), (\ref{eq_e}), and the above notation,
the power market model can be compactly expressed as
\begin{equation}\label{eq_power_market}
    \bm{\dot{x}}(t)=\bm{A} \bm{x}(t)+\bm{b}+\bm{\tau}\lambda(t)+\bm{Bw}(t).
\end{equation}

For the case where $\bm{w}(t)=0$, i.e., no uncertainty and no power input from the RES,
the power model (\ref{eq_power_market}) is reduced to the scenario considered in \cite{Alva0,Alva1,Alva2}.
To balance the energy,
the price can be controlled by the differential equation
\begin{equation}\label{eq_lambda}
    \dot{\lambda}(t)=\frac{-e(t)}{\tau_\lambda}
\end{equation}
where $\tau_\lambda$ is a speed constant that has the same role as $\tau_g$ and $\tau_d$.
The pricing scheme in (\ref{eq_lambda}) is referred to as
area control error (ACE) pricing scheme \cite{Alva1}, which depends on the feedback of real time energy imbalance.
In the language of control theory, the ACE pricing scheme is a dynamic pricing controller as it involves the price dynamics.
For the case in which $\bm{w}(t)\not =0$, a robust pricing scheme against the disturbances $\bm{w}(t)$ is needed.
The next section is dedicated to designing  $\lambda(t)$ for management of $e(t)$ in the presence of $\bm{w}(t)\not =0$.

\begin{rmk}\label{rmk_market_power_sys}
In general, deploying price-based controllers, e.g., the ACE and the proposed pricing schemes,
in a real-world scenario requires using additional knowledge extracted from the underlying power systems.
For instance, if
a power system uses
synchronous machines modeled by a 3rd order flux decay model or a 4th order two axis model \cite{pow_sys},
knowledge of ``average frequency deviation''
from the machine needs to be
 added into the market dynamics as a substantial measurement of imbalanced energy \cite{interco}.
However, such knowledge depends on explicit power system structures and the corresponding mathematical formulations are beyond the scope of this paper.
We refer the reader  to \cite{Alva2} and \cite{interco}
for relevant discussions about the interconnection of power systems and market dynamics.
In \cite{Alva2}, an automatic voltage regulator model interconnected with market dynamics was
examined.
In  \cite{interco},
such interconnection was further studied by using
the New England 39 bus test system, including generator/turbine/governor dynamics.
\end{rmk}

\section{Proposed Robust Pricing Scheme}\label{sec_proposed}

In this section, a fuzzy system is proposed to replace the power market defined in (\ref{eq_power_market})
so that a robust pricing scheme can be constructed based on it.
We will consider an $H_{\infty}$ design for the proposed scheme due to its robustness against system disturbances, such as the uncertainty $\Delta_g(t)$ and $\Delta_d(t)$, and the fluctuating power input $in(t)$ that is contained in $\bm{w}(t)$.
Unlike  the ACE pricing scheme (\ref{eq_lambda}) which  employs only the information of $e(t)$, the proposed pricing scheme utilizes the feedback of power supply $p_g(t)$, power demand $p_d(t)$, and imbalance energy $e(t)$.

To facilitate the design, our strategy is to interpolate $\bm{A} \bm{x}(t)+\bm{b}$ in different operating regions by
several linear systems of the form $\bm{A}_m \bm{x}(t)$.
That is, referring to (\ref{eq_power_market}), the system
\begin{equation}\label{eq_be_approx}
   \bm{y}(t)=\bm{A} \bm{x}(t)+\bm{b}
\end{equation}
is represented by  \cite{sys_id1,sys_id2}
\begin{equation}\label{eq_fuzzy_sys}
\begin{split}
  \mbox{Rule } &\; m \\
     \mbox{If } & p_g(t) \mbox{ is } F_{m1},p_d(t) \mbox{ is } F_{m2}, \mbox{and } e(t) \mbox{ is } F_{m3} \\
   \mbox{Then }   &   \bm{y}(t)=\bm{A}_m \bm{x}(t)
\end{split}
\end{equation}
for $m=1,2,...,M$, where $M$ represents the number of fuzzy rules. The premises in the fuzzy system
(\ref{eq_fuzzy_sys}) are the states  $p_g(t),p_d(t)$, and $e(t)$. $F_{m1},F_{m2}$, and $F_{m3}$ are fuzzy membership functions.
According to (\ref{eq_fuzzy_sys}), system (\ref{eq_be_approx}) can be represented by a fuzzy system as
\begin{equation}\label{eq_be_approx2}
   \bm{y}(t)= \sum_{m=1}^M  h_m(\bm{x}(t))\bm{A}_m \bm{x}(t)+\Delta_{\bm{x}}
\end{equation}
where
\begin{equation}\label{eq_fuzzy_base}
h_m(\bm{x}(t))=
\frac{ F_{m1}(p_g(t)) F_{m2}(p_d(t)) F_{m3}(e(t)) }{ \sum_{m'=1}^M F_{m'1}(p_g(t)) F_{m'2}(p_d(t)) F_{m'3}(e(t))}.
\end{equation}
The term
\begin{equation*}
  \Delta_{\bm{x}}= (\bm{A} \bm{x}(t)+\bm{b} ) -  \sum_{m=1}^M  h_m(\bm{x}(t))\bm{A}_m \bm{x}(t)
\end{equation*}
denotes the approximation error, which can be very small if sufficient fuzzy rules are used.
 The approximation error  $\Delta_{\bm{x}}$ is omitted in the ensuing derivation by assuming that a large value of $M$ is employed.
Each $F_{mn}$ can be interpreted as the set to which a certain premise belongs with degree $F_{mn}(\cdot)$.
Therefore, $F_{mn}(\cdot)$ is always non-negative and, according to (\ref{eq_fuzzy_base}), we have $h_m(\bm{x}(t))\geq 0$
with $\sum_{m=1}^M h_m(\bm{x}(t))=1$.
 For example,
$p_g(t)$ belongs to $F_{m1}$  with the degree $F_{m1}(p_g(t))$.
 In our simulations, we will show the construction of $F_{mn}$.
Once the membership functions $F_{mn}$ are assigned, the matrices $\bm{A}_m,m=1,2,...,M$ can be evaluated by least-squares methods. At this point, we assume that $F_{mn}$ and $\bm{A}_m$ are available for further manipulation.

Similarly to the fuzzy system (\ref{eq_fuzzy_sys}), the proposed pricing  scheme is also constructed by fuzzy rules as
\begin{equation}\label{eq_fuzzy_u}
\begin{split}
  \mbox{Rule } &\; m \\
     \mbox{If } & p_g(t) \mbox{ is } F_{m1},p_d(t) \mbox{ is } F_{m2}, \mbox{and } e(t) \mbox{ is } F_{m3} \\
   \mbox{Then }   &   \lambda(t)=\bm{K}_m \bm{x}(t)
\end{split}
\end{equation}
for $m=1,2,...,M$, where $\bm{K}_m$ represents the control gain to be designed.
 According to (\ref{eq_fuzzy_u}), the overall pricing scheme
can be obtained as
\begin{equation}\label{eq_over_u}
   \lambda(t)= \sum_{m=1}^M  h_m(\bm{x}(t))\bm{K}_m \bm{x}(t)
\end{equation}
where the fuzzy basis $ h_m(\bm{x}(t))$ is defined in (\ref{eq_fuzzy_base}).
The same premises and membership functions as in (\ref{eq_fuzzy_sys}) are adopted in (\ref{eq_fuzzy_u})
for further integration. In contrast to the ACE pricing scheme (\ref{eq_lambda}), the proposed scheme (\ref{eq_over_u})
is a static pricing controller as it does not involve the price dynamics \cite{Boyd_LMI}.

Based on (\ref{eq_be_approx}), (\ref{eq_be_approx2}), and (\ref{eq_over_u}),
the power market system in (\ref{eq_power_market}) can be equivalently expressed as
\begin{equation}\label{eq_equi}
    \begin{split}
       \bm{\dot{x}}(t){ } ={ } &  \bm{y}(t)+\bm{\tau}\lambda(t)+\bm{Bw}(t)\\
       { } ={ }  &  \sum_{m=1}^M   h_m(\bm{x}(t)) ( \bm{A}_m+  \bm{\tau}  \bm{K}_m ) \bm{x}(t)+\bm{Bw}(t)\\
{ } :={ } &  \sum_{m=1}^M   h_m(\bm{x}(t)) \bm{\tilde{A}}_m \bm{x}(t)+\bm{Bw}(t)
    \end{split}
\end{equation}
where $\bm{\tilde{A}}_m=\bm{A}_m+  \bm{\tau}  \bm{K}_m$.
The aim of the pricing scheme design is to drive $e(t)$ as close to zero  as possible.
This can be done by choosing an appropriate system output $\bm{z}(t)$ and control gains $\bm{K}_m $ such that
 the $H_\infty$ performance criterion \cite{Boyd_LMI}
\begin{equation}\label{eq_Hinf}
    \int_0^\infty   \bm{z}(t)^T\bm{z}(t)  -\gamma^2 \bm{w}(t)^T\bm{w}(t) dt <0
\end{equation}
is satisfied.
The physical meaning of  (\ref{eq_Hinf}) is described as follows:
the energy of $\bm{z}(t)$ is controlled against the energy of disturbance $\bm{w}(t)$ so that a prescribed
$H_\infty$ attenuation level $\gamma >0$ is guaranteed. For a better understanding, let us suppose that $\bm{w}(t)\not=0$ and $\bm{z}(t),\bm{w}(t)\in \mathfrak{L}_2[0\;\infty)$, i.e.,
\begin{equation}\label{eq_def_L2}
     \int_0^\infty \bm{z}(t)^T\bm{z}(t) dt < \infty,  \int_0^\infty \bm{w}(t)^T\bm{w}(t) dt < \infty.
\end{equation}
The condition (\ref{eq_Hinf}) is then equivalent to
\begin{equation}\label{eq_Hinf2}
 \sup_{\bm{w}(t)\not=0}  \frac{ \sqrt{\int_0^\infty   \bm{z}(t)^T\bm{z}(t) dt}}{ \sqrt{\int_0^\infty  \bm{w}(t)^T\bm{w}(t) dt}}  <\gamma
\end{equation}
which explains why $\gamma$ is regarded as an attenuation level.

If we  choose
\begin{equation}\label{eq_output}
\bm{z}(t)=
\left[
  \begin{array}{c}
    e(t) \\
    \varepsilon \lambda(t)\\
  \end{array}
\right]
\end{equation}
for a small $\varepsilon>0$,
then (\ref{eq_Hinf2}) can be interpreted as follows:
 the imbalanced energy $e(t)$ against
the disturbances $\bm{w}(t)$  is mainly controlled such that
 the attenuation ratio is less than the prescribed level $\gamma$.
The price $\lambda(t)$ pre-multiplied by $ \varepsilon$  as shown in (\ref{eq_output})
is also involved in the output $\bm{z}(t)$ because a  large price perturbation can be undesirable
in practice. By setting a small value of $\varepsilon$,
the imbalanced energy $e(t)$ can be controlled by an appropriate price perturbation.
For simplicity, we define
\begin{equation}\label{eq_def_C}
\bm{C}=
\left[
  \begin{array}{ccc}
    0 & 0 & 1 \\
    0 & 0 & 0 \\
  \end{array}
\right] \mbox{ and }
\bm{D}=
\left[
  \begin{array}{c}
    0 \\
    \varepsilon   \\
  \end{array}
\right].
\end{equation}
According to (\ref{eq_over_u}) and (\ref{eq_def_C}), the system output $\bm{z}(t)$ in (\ref{eq_output}) can then be
expressed as
\begin{equation}\label{eq_output2}
\begin{split}
\bm{z}(t)= \;& \bm{C}\bm{x}(t)+\bm{D}\lambda(t) \\
   =  \;&  \sum_{m=1}^M  h_m(\bm{x}(t))  [ \bm{C}+\bm{D}\bm{K}_m ] \bm{x}(t)\\
:= \;& \sum_{m=1}^M  h_m(\bm{x}(t)) \bm{\tilde{C}}_m \bm{x}(t)
\end{split}
\end{equation}
where  $\bm{\tilde{C}}_m=  \bm{C}+\bm{D}\bm{K}_m $.

To guarantee the condition in (\ref{eq_Hinf}),
 let us consider the quadratic Lyapunov function \cite{Lyap}
 \begin{equation}\label{eq_Lyapu}
    V(\bm{x})=\bm{x}(t)^T\bm{P}\bm{x}(t)
 \end{equation}
for some positive matrix $\bm{P}\succ 0$ to be determined.
The $H_\infty$ performance criterion (\ref{eq_Hinf}) is satisfied if
\begin{equation}\label{eq_suff}
\dot{V}(\bm{x})+\bm{z}(t)^T\bm{z}(t)-\gamma^2 \bm{w}(t)^T\bm{w}(t)<0
\end{equation}
holds true \cite{Boyd_LMI}. Note that  $\dot{V}(\bm{x})=2\bm{\dot{x}}(t)^T\bm{P}\bm{x}(t)$.
By substituting (\ref{eq_equi}) and (\ref{eq_output2}) into the left-hand side of (\ref{eq_suff}),
we have
\begin{equation}\label{eq_proof}
    \begin{split}
      &  \dot{V}(\bm{x})+\bm{z}(t)^T\bm{z}(t)-\gamma^2 \bm{w}(t)^T\bm{w}(t) \\
 \leq{ }  &  2 [  \sum_{m=1}^M   h_m(\bm{x}(t)) \bm{\tilde{A}}_m \bm{x}(t)+\bm{Bw}(t) ]^T\bm{P}\bm{x}(t)\\
   &    +   \sum_{m=1}^M   h_m(\bm{x}(t)) \bm{x}(t)^T \bm{\tilde{C}}_m^T   \bm{\tilde{C}}_{m}  \bm{x}(t)  - \bm{w}(t)^T   ( \gamma^2 \bm{I}_3 ) \bm{w}(t) \\
  ={ }  &
\sum_{m=1}^M   h_m(\bm{x}(t))
  \left[
    \begin{array}{c}
     \bm{x}(t) \\
     \bm{w}(t) \\
    \end{array}
  \right]^T
 \left\{
   \begin{array}{c}
  \left[
    \begin{array}{cc}
     \bm{\tilde{A}}_m^T \bm{P}+\bm{P}\bm{\tilde{A}}_m & \bm{PB} \\
      \star & - \gamma^2 \bm{I}_3 \\
    \end{array}
  \right]
     \\
   \end{array}
 \right.
\\
  &
\left.
 +
   \left[
    \begin{array}{c}
    \bm{\tilde{C}}_m^T  \\
     \bm{0} \\
    \end{array}
  \right]
  \bm{I}_2^{-1}
  \begin{array}{c}
  \left[
    \begin{array}{cc}
  \bm{\tilde{C}}_m  & \bm{0} \\
    \end{array}
  \right]
  \end{array}
\right\}
  \left[
    \begin{array}{c}
     \bm{x}(t) \\
     \bm{w}(t) \\
    \end{array}
  \right] \\
:={ }&
\sum_{m=1}^M   h_m(\bm{x}(t))
  \left[
    \begin{array}{c}
     \bm{x}(t) \\
     \bm{w}(t) \\
    \end{array}
  \right]^T
\bm{\Phi}
  \left[
    \begin{array}{c}
     \bm{x}(t) \\
     \bm{w}(t) \\
    \end{array}
  \right].
    \end{split}
\end{equation}
The inequality comes from the fact that (Lemma 2 \cite{Hinf})
\begin{equation*}
\begin{split}
   & \sum_{m=1}^M   h_m(\bm{x}(t)) \bm{x}(t)^T \bm{\tilde{C}}_m^T   \sum_{m'=1}^M   h_{m'}(\bm{x}(t)) \bm{\tilde{C}}_{m'}  \bm{x}(t)\\
 \leq \;   &  \sum_{m=1}^M   h_m(\bm{x}(t)) \bm{x}(t)^T \bm{\tilde{C}}_m^T    \bm{\tilde{C}}_{m}  \bm{x}(t).
\end{split}
\end{equation*}
In (\ref{eq_proof}), the mark ``$\star$'' denotes symmetric terms of a symmetric matrix. i.e., $(\bm{PB})^T$ in this case.

Based on (\ref{eq_proof}), a sufficient condition for the validity of (\ref{eq_suff}) is $\bm{\Phi}\prec 0$, which
is equivalent to (Schur complement \cite{Boyd_conv,Boyd_LMI})
\begin{equation}\label{eq_LMI1}
  \left[
      \begin{array}{ccc}
        \bm{\tilde{A}}_m^T \bm{P}+  \bm{P}\bm{\tilde{A}}_m & \bm{PB}  & \bm{\tilde{C}}_m ^T \\
        \star & -\gamma^2 \bm{I}_3 & \bm{0} \\
        \star & \star & -\bm{I}_2 \\
      \end{array}
  \right]\prec 0
\end{equation}
for $m=1,2,...,M$.
After pre-multiplying and post-multiplying (\ref{eq_LMI1}) by $diag(\bm{P}^{-1},\bm{I}_3,\bm{I}_2)$,
we have
\begin{equation}\label{eq_LMI2}
  \left[
      \begin{array}{ccc}
        \bm{Q} \bm{\tilde{A}}_m^T +  \bm{\tilde{A}}_m \bm{Q} & \bm{B}  & \bm{Q} \bm{\tilde{C}}_m ^T \\
        \star & -\gamma^2 \bm{I}_3 & \bm{0} \\
        \star & \star & -\bm{I}_2 \\
      \end{array}
  \right]\prec 0,\; \forall m
\end{equation}
where
\begin{equation}\label{eq_Q}
     \bm{Q}=\bm{P}^{-1}\succ 0.
\end{equation}

In (\ref{eq_LMI2}),   $\bm{K}_m$ is contained in $\bm{\tilde{A}}_m$ and $\bm{\tilde{C}}_m$.  As   $\bm{Q}$ and  $\bm{K}_m$ are variables and coupled, the matrix inequality (\ref{eq_LMI2}) is not linear.
It is essential to have an LMI since such is convex and hence, can be efficiently solved.
To obtain a feasible solution of (\ref{eq_LMI2}), let us define
\begin{equation}\label{eq_change_var}
    \bm{Y}_m=\bm{K}_m \bm{Q}.
\end{equation}
By substituting (\ref{eq_change_var}) into the terms $ \bm{Q} \bm{\tilde{A}}_m^T +  \bm{\tilde{A}}_m \bm{Q}$ and
$\bm{Q} \bm{\tilde{C}}_m ^T$ in (\ref{eq_LMI2}), we have
\begin{equation}\label{eq_LMI3}
\begin{split}
   &
\left[
      \begin{array}{ccc}
       \bm{A}_m \bm{Q} + \bm{\tau}\bm{Y}_m + (\star)& \bm{B}  & \bm{Q}\bm{C}^T+ \bm{Y}_m^T \bm{D}^T\\
        \star & -\gamma^2 \bm{I}_3 & \bm{0} \\
        \star & \star & -\bm{I}_2 \\
      \end{array}
  \right]\prec 0
 \\
    & \mbox{for all } m, \mbox{ and }\bm{Q}\succ 0
\end{split}
\end{equation}
where $(\star)$ represents $( \bm{A}_m \bm{Q} + \bm{\tau}\bm{Y}_m)^T$.
The matrix inequality (\ref{eq_LMI3}) is an LMI in $\bm{Q}\succ 0$ and $\bm{Y}_m$.
Referring to (\ref{eq_change_var}), the control gains $\bm{K}_m$ can be obtained by
$\bm{K}_m=\bm{Y}_m \bm{Q}^{-1}$ such that the $H_\infty$ performance in (\ref{eq_Hinf}) holds true.

We summarize the proposed pricing scheme as follows.
For the market power system in (\ref{eq_power_market}), a robust pricing scheme is proposed in the form of (\ref{eq_over_u}),
where the control gains $\bm{K}_m$ are evaluated by solving (\ref{eq_LMI3}).
The LMI (\ref{eq_LMI3}) can be solved by successively lowering the
value of $\gamma$ until (\ref{eq_LMI3}) becomes infeasible. The smallest $\gamma>0$ that guarantees
the feasibility of (\ref{eq_LMI3}) can be used and the corresponding $\bm{Q}$ and $\bm{Y}_m$
can be obtained to further evaluate $\bm{K}_m$ \cite{Boyd_conv,Boyd_LMI}.
In our simulations, the proposed scheme is compared to the ACE pricing scheme presented by the differential equation in (\ref{eq_lambda}).

\begin{rmk}\label{rmk_storage}
A microgrid often possesses storage capabilities and requires
 the stored energy $e(t)$ to be maintained at a certain energy level to  facilitate both ordinary and emergency power use \cite{ESS_Da}.
For a microgrid with an energy storage system, it becomes more reasonable to consider $e(t)\rightarrow q>0$, where $q$ represents the desired energy level.
This aspect can be included in our proposed scheme by using a change of variables, i.e., $\tilde{e}(t)=e(t)-q$, as shown in the following.
In (\ref{eq_p_g}), the feedback term $k e(t)$ is replaced by   $k \tilde{e}(t)$ because the additional cost is now introduced
by not achieving the desired energy level $q$, i.e., $\tilde{e}(t)\not=0$.
Since $\tilde{e}(t)$ is different from $e(t)$ by a constant term, they have the same dynamics $\dot{\tilde{e}}(t)=\dot{e}(t)$
as expressed in (\ref{eq_e}). For the premise variable in the fuzzy rules, the augmented state $\bm{x}(t)$ in (\ref{eq_over_u}) and
the system output $\bm{z}(t)$ in (\ref{eq_output}), $\tilde{e}(t)$ replaces the role of
 $e(t)$.
It can be found that the change of variables results in the same LMI constraint in (\ref{eq_LMI3}), while
the only difference is the interpretation of ``imbalanced energy''.
In such a configuration,
the energy is imbalanced if the stored energy $e(t)$ is not maintained at a desired working level $q$ or, equivalently,
$\tilde{e}(t)\not =0$.
\end{rmk}

\begin{rmk}\label{rmk_network_micro}
When a network of microgrids is equipped with an energy management system (EMS), our proposed scheme
turns into a centralized design after suitable modification.
For this centralized configuration,
each microgrid may be connected to another microgrid so that
energy state information of microgrids is collected and used to achieve certain network performance objectives \cite{EMS1}.
In this case, the EMS functions from a whole networked system perspective.
However, the network size should be reasonable for efficient energy management, and
the proposed pricing scheme needs further modification to include the interactive relation between microgrids.
In contrast,
the proposed pricing scheme is readily applicable to a decentralized configuration
when a microgrid is only connected to the conventional power grid and the EMS
 operates  within the scope of a microgrid.
\end{rmk}

\begin{rmk}
There exist forecasting techniques \cite{forecast1,forecast2} that are able to provide good predictions of the power input $in(t)$ given by RESs.
When these prediction techniques are employed by the proposed scheme, $in(t)$  can be modeled as $in(t)=\widehat{in}+\Delta_{in}(t)$, where $\widehat{in}$ and $\Delta_{in}(t)$
represent the predicted average power input and the prediction error, respectively.
The column vectors $\bm{b}$ and $\bm{w}(t)$ in (\ref{eq_power_market})
should be modified as
\begin{equation*}
 \bm{b}=
 \left[
   \begin{array}{ccc}
    - \frac{\hat{b}_g}{\tau_g} & \frac{\hat{b}_d}{\tau_d} & \widehat{in} \\
   \end{array}
 \right]^T
\mbox{ and }
\bm{w}(t)=
 [\Delta_g(t)\;\Delta_d(t)\;\Delta_{in}(t)]^T
\end{equation*}
respectively.
The knowledge on $\widehat{in}$ is then updated over time.
When $\widehat{in}$ is updated, $\bm{A}_m,m=1,2,...,M$ and thus $\bm{K}_m,m=1,2,...,M$ need to be re-evaluated as well.
\end{rmk}

\section{Numerical Examples}\label{sec_sim}

In this section, we describe simulations of the power market behavior according to its dynamics in (\ref{eq_power_market}).
TABLE \ref{tab_par} lists the numerical values of the system parameters used in these simulations.
Two numerical examples are considered. The first example considers
the market behavior without system uncertainty and  power input from the RES, i.e., $\bm{w}(t)=0$. The second example extends to the case where $\bm{w}(t)\not=0$, i.e., the power market behavior for a microgrid is investigated.

\begin{table}
  \centering
  \caption{Power Market Parameters}\label{tab_par}
\begin{tabular}{|cc|cc|}
  \hline
  $c_g$ & 0.4 &    $c_d$ & 0.5   \\
 $\tau_g$ & 0.2   & $\tau_d$ & 0.25 \\
 $\hat{b}_g$ & 2 & $\hat{b}_d$ & 10 \\
$\tau_\lambda$ & 100  &   $k$ & 0.1 \\
  $\lambda(0)$ & 4.66 &  $e(0)$  &  0\\
 $p_g(0)$ & 10.4 &   $p_d(0)$ & 13\\
  \hline
\end{tabular}
\end{table}

\begin{figure}
  \centering
  \includegraphics[width=8cm]{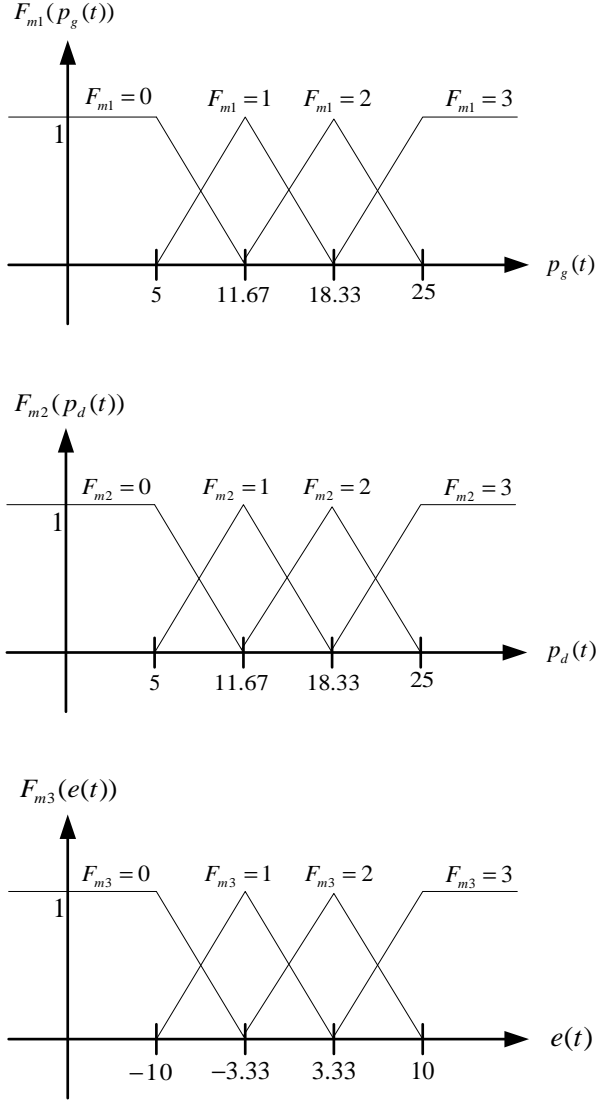}\\
  \caption{Four fuzzy membership functions denoted by $F_{mn}=0,F_{mn}=1,F_{mn}=2$ and $F_{mn}=3$.
The set $[5,25]\times[5,25]\times[-10,10]$ is the input space for the fuzzy systems (\ref{eq_fuzzy_sys}) and (\ref{eq_fuzzy_u}).
Each input range is uniformly partitioned by fuzzy membership functions.
}\label{fig_Fmn}
\end{figure}

For the proposed pricing scheme in (\ref{eq_over_u}),
the fuzzy membership functions $F_{mn}$ in (\ref{eq_fuzzy_sys}) and (\ref{eq_fuzzy_u})  were constructed according to Fig. \ref{fig_Fmn}.
The input ranges $[5, 25],[5,25]$ and $[-10,10]$ were considered for the premises $p_g(t)$, $p_d(t)$ and $e(t)$, respectively.
As the range of each premise $p_g(t)$, $p_d(t)$ or $e(t)$ in fuzzy rules is covered  by four membership functions,
denoted by $F_{mn}=0, F_{mn}=1, F_{mn}=2$ and  $F_{mn}=3$,
there are $M=4^3=64$ fuzzy rules. For instance,  $(F_{m1},F_{m2},F_{m3})=(0,2,3)$ is one of the $64$ fuzzy rules.
 These fuzzy membership functions in Fig. \ref{fig_Fmn} were adopted because of their simplicity. Another popular choice of
$F_{mn}$ is a bell-shaped function \cite{sys_id2}.
For  more sophisticated fuzzy schemes that use fewer fuzzy rules, the reader can refer to \cite{Hinf,sys_id1,sys_id2}  and the references therein.

Take the input premises $(p_g(t),p_d(t),e(t))=(11.67,8.335,-1)$ and the fuzzy rule $(F_{m1},F_{m2},F_{m3})=(1,0,3)$ as an example.
Referring to Fig. \ref{fig_Fmn} with  $(F_{m1},F_{m2},F_{m3})=(1,0,3)$ , we have
 \begin{equation*}
{\small
\begin{split}
 F_{m1}(p_g(t)){ } = { }& \max\{ \min \{   \frac{1}{11.67-5}  (p_g(t)-5),  \\
 &    \frac{1}{18.33-11.67}  (18.33-p_g(t))\},0\} \\
F_{m2}(p_d(t)){ }  ={ } &   \max\{ \min \{ \frac{1}{11.67-5}  (11.67-p_d(t)), 1 \},0\} \\
 F_{m3}(e(t)){ }  = { }  &  \max\{ \min \{ \frac{1}{10-3.33}  (e(t)-3.33), 1 \},0\}
\end{split}
}
 \end{equation*}
implying that
$F_{m1}(11.67)=1$, $F_{m2}(8.335)=0.5$ and $F_{m3}(-1)=0$.
 This can be interpreted as follow. The premises $p_g(t)$, $p_d(t)$ and $e(t)$ belong to
the fuzzy membership functions $F_{m1}=1$, $F_{m2}=0$ and $F_{m3}=3$ with the degrees  $1, 0.5$ and $0$, respectively.
The term ``belong to'' is used because a fuzzy membership function is often referred to as a fuzzy set.
In this case, $h_m([11.67\;8.335\;-1]^T)$ represents the degree of fulfillment of the $m$th fuzzy rule, where $h_m$ is defined in (\ref{eq_fuzzy_base}).

\begin{figure*}
 \centering
\subfigure[] {
\label{price_ace}
\includegraphics[width=8.3cm]{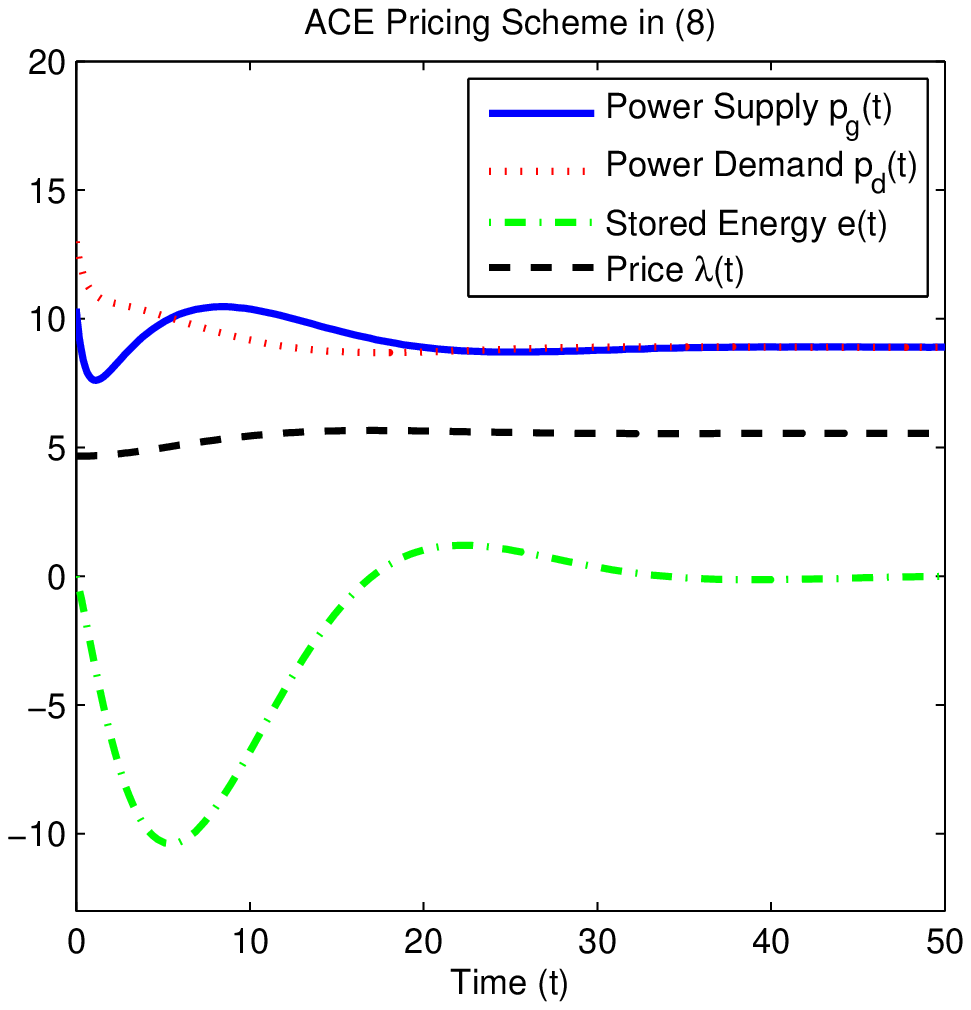}
}
\subfigure[] {
\includegraphics[width=8.3cm]{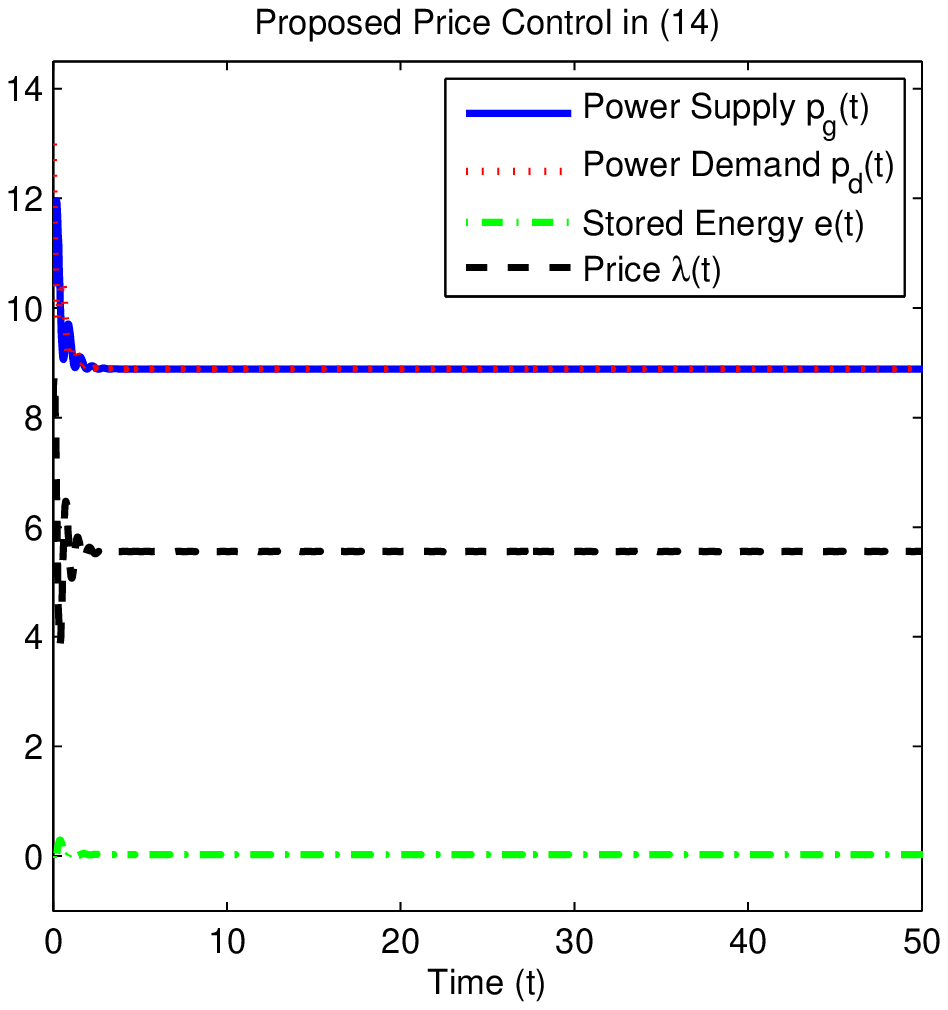}
}
\caption{Traditional power market in Example 1 with $\bm{w}(t)=0$:
(a) The ACE pricing in (\ref{eq_lambda}); (b) The proposed robust pricing scheme in (\ref{eq_over_u}).
}\label{fig_w}
\end{figure*}

\begin{figure*}
 \centering
\subfigure[] {
\label{price_ace}
\includegraphics[width=8.3cm]{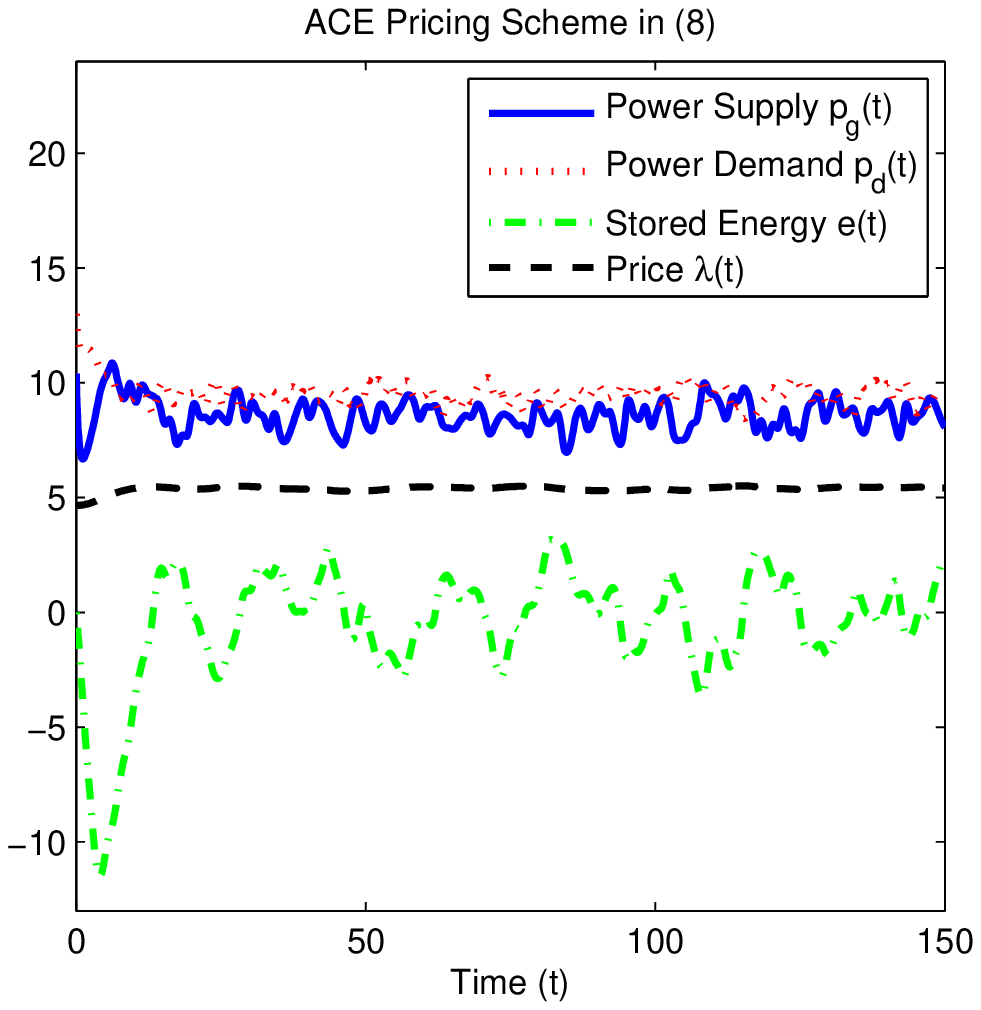}
}
\subfigure[] {
\includegraphics[width=8.3cm]{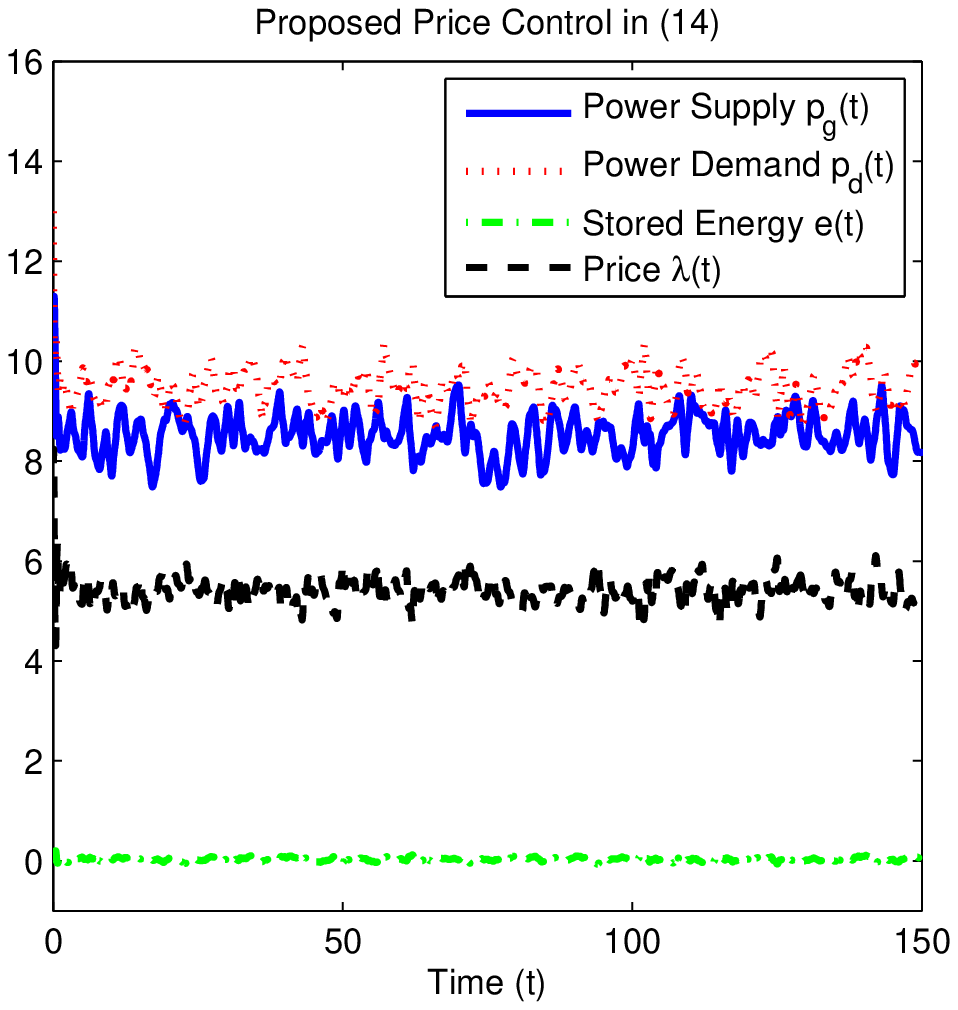}
}
\caption{
Power market for a microgrid in Example 2 with $\bm{w}(t)$ defined by (\ref{eq_wt1}):
(a) The ACE pricing in (\ref{eq_lambda}); (b) The proposed robust pricing scheme  in (\ref{eq_over_u}).
}\label{fig_w2}
\end{figure*}

 Once the fuzzy rules have been constructed, i.e., fuzzy membership functions are assigned to each rule, the matrices $\bm{A}_m$ in (\ref{eq_fuzzy_sys}) can be obtained as follows:
first, a sequence of input vectors $\bm{x}_\ell,\ell=1,2,...,L,$ are randomly generated
from the input space  $[5,25]\times[5,25]\times[-10,10]$. The number $L=1500$ was chosen.
A sequence of output vectors $\bm{y}_\ell,\ell=1,2,...,L,$ can be obtained by inputting  $\bm{x}_\ell$ into (\ref{eq_be_approx}).
 By substituting $\bm{x}_\ell$ and $\bm{y}_\ell$ into
(\ref{eq_be_approx2}), we have $3L$ equations with $\bm{A}_m,m=1,2,...,M,$ as variables to be solved.
The matrices $\bm{A}_m$ can be estimated by using least-squares methods.
For $M=64$ fuzzy rules, the approximation error $\Delta_{\bm{x}}$ in (\ref{eq_be_approx2}) is relatively small with respect to the input energy, i.e.,
\begin{equation*}
   \sup_{\bm{x}_\ell \not=0,\ell=1,2,...,L}  \frac{\Delta_{\bm{x}_\ell}^T \Delta_{\bm{x}_\ell}   }{\bm{x}_\ell^T \bm{x}_\ell}=0.0193.
\end{equation*}
To obtain the gain matrices $\bm{K}_m$ in the proposed pricing scheme,
$\bm{A}_m$ for $m=1,2,...,M$ were substituted into (\ref{eq_LMI3}).
For prescribed values of $\gamma^2=2$ in (\ref{eq_LMI3})
and $\varepsilon=0.1$ in (\ref{eq_def_C}),
(\ref{eq_LMI3}) is an LMI in $\bm{Q}$ and $\bm{Y}_m$. As an LMI problem is a convex problem, (\ref{eq_LMI3}) with $\bm{Q}\succ0$ can be efficiently
solved using existing algorithms such as interior-point methods \cite{Boyd_conv,Boyd_LMI}.

\subsection{Example 1: $\bm{w}(t)=0$}

For the first simulation example, the case where $\bm{w}(t)=0$ was considered, i.e., no uncertainty in the marginal cost ($\hat{b}_g=b_g$) and the marginal benefit ($\hat{b}_d=b_d$), and no power input from the RES ($in(t)=0$). This
is the scenario for which  the ACE pricing scheme (\ref{eq_lambda}) was designed \cite{Alva1}.
We have compared the proposed pricing scheme (\ref{eq_over_u}) to the ACE pricing scheme.
The initial conditions $\lambda(0),e(0),p_g(0)$ and $p_d(0)$ were listed in TABLE
\ref{tab_par} ($\lambda(0)$ is needed in the ACE pricing scheme).
The system behavior was observed from time $t=0$ to $t =50$.
$(p_g,p_d,e,\lambda)=(8.89, 8.89, 0, 5.56)$ is
the equilibrium point of the augmented system (\ref{eq_power_market}) and (\ref{eq_lambda}) with $\bm{w}(t)=0$.

In this example, the energy is balanced if $p_g(t)$ and $p_d(t)$ converge to the same value. See Figs. \ref{fig_w}(a) and (b) for the resulting performance.
Although both pricing schemes can stabilize the power market system, the proposed approach reaches the steady state
more quickly than the ACE pricing scheme,
the imbalanced energy $e(t)$ in particular.
As shown in Fig. \ref{fig_w}(b),
 the superior energy imbalance management
of  the proposed scheme results from vibrating
the price so that $e(t)$ could converge to zero rapidly. In contrast, the ACE scheme has less price vibration and a slower
 convergence rate of the energy imbalance as shown in Fig. \ref{fig_w}(a).
It is interesting to notice that although the proposed pricing mechanism is different from the ACE pricing scheme,
it still converges to the equilibrium point  $(p_g,p_d,e,\lambda)=(8.89,8.89,0,5.56)$ of ACE pricing.

\subsection{Example 2: $\bm{w}(t)\not =0$}

For the second example,
let us consider a power market system for a microgrid, i.e., $\bm{w}(t)\not= 0$.
The extra power from the RES, and the natural uncertainty in marginal cost and benefit \cite{Alva1}
were involved as the overall system disturbances.
In this case, the overall disturbances $\bm{w}(t)$ in (\ref{eq_power_market}) were simulated by
\begin{equation}\label{eq_wt1}
   \bm{w}(t)=
{\small
\left[
  \begin{array}{ccc}
 rand_{[-0.5,0.5]}(t)  & rand_{[-0.4,0.6]}(t) & rand_{[0,2]}(t) \\
  \end{array}
\right]^T}
\end{equation}
where $rand_{[q_1,q_2]}(t)$ represents a random process that is uniformly distributed over the range $[q_1,q_2]$.
As the values $in(t)$ can assume
are always positive, $in(t)=rand_{[0,2]}(t)$ in (\ref{eq_wt1}) was employed to  model the input power energy to the microgrid.

In this example, the power supply $p_g(t)$ must be less than the power demand $p_d(t)$ in order to avoid energy imbalance because
the extra power input $in(t)>0$ exists. To obtain a clear view of system trajectories, the power market behavior was examined
from time $t=0$ to $t=150$.
Figs. \ref{fig_w2}(a) and (b) show that the proposed pricing scheme outperforms the ACE scheme
through superior energy imbalance management.
As expected, we see that $p_g(t)< p_d(t)$ in the proposed scheme according to Fig. \ref{fig_w2}(b).
 The price vibration is used to robustly stabilize $e(t)$ against the fluctuating power input $in(t)$ and the system uncertainty. In contrast,
the ACE pricing scheme in Fig. \ref{fig_w2}(a) involves little price vibration but results in severe vibration of  $e(t)$.
In this example, the four states $p_g(t),p_d(t),e(t)$ and $\lambda(t)$ of both pricing schemes
fluctuate around
the same equilibrium $(p_g,p_d,e,\lambda)=(8.89,8.89,0,5.56)$.

\begin{rmk}
Suppose that the scenarios covered by Remarks \ref{rmk_storage} and \ref{rmk_network_micro} are considered in our scheme.
The generated power $p_g(t)$ is then consumed by $N$ microgrids with $N>1$. Let  $p_{d_n}(t)$ denote the power demand of microgrid $n$.
The corresponding stored energy $e_n(t)$ is required to approach a certain energy level $q_n>0$ for $n=1,2,...,N$.
Referring to Figs. \ref{fig_w} and \ref{fig_w2}, $p_g(t)$ is mostly less than or equal to $p_d(t)$, and $e(t)$ vibrates around zero in our simulations. In contrast, when the scenarios are considered,
$p_g(t)$ should become larger than  $p_{d_n}(t)$ provided that the RESs only offer a small amount of power inputs to the microgrids. The relation $p_g(t)>p_{d_n}(t)$  results from the existence of multiple microgrids  such that $N$ microgrids need to share the generated power $p_g(t)$. The difference  $p_g(t)-p_{d_n}(t)$ can be enlarged upon increasing
$N$. In this case, $e_n(t)$ should vibrate around $q_n>0$ instead of zero.
\end{rmk}

From the previous two examples, it was found that price vibration was a crucial factor for energy imbalance management.
In the first example, the vibration in the proposed pricing scheme occurred only at the beginning and was alleviated over the remaining time. The initial vibration was used to deal with the imbalanced initial conditions. In the second example, the price vibrated  continuously due to the existing disturbances, i.e., the fluctuating power input and the uncertainty in marginal cost/benefit. By comparing two different pricing schemes, we can summarize that
the proposed pricing scheme is more robust against  disturbances than the ACE scheme.
This is mainly because  the proposed pricing scheme  fully utilizes all states, i.e, $p_g(t),p_d(t)$ and $e(t)$,
while the ACE scheme employs only the feedback of $e(t)$.
From the performance of the proposed methodology, we conclude that, by adjusting the market price appropriately, it is possible to robustly balance the energy against uncertainty in marginal cost/benefit and the fluctuating power input from the RES.

\section{Conclusion}\label{sec_con}

This paper has considered  power market behavior in a microgrid, which is different from the traditional power market model.
The market dynamics studied can be regarded as a generalization of the traditional market dynamics.
A novel pricing scheme for the energy management in a microgrid has been proposed. The underlying idea is to use fuzzy systems
 together with an LMI approach to assure the robustness of market dynamics.

 In the language of control theory, the proposed pricing scheme
is a static pricing controller, while the existing ACE scheme is a dynamic pricing controller as it employs the price dynamics
for energy imbalance management.
We do not conclude that a static pricing scheme is better than a dynamic pricing scheme. In fact, dynamic pricing is of course a generalization of static pricing. We have intended to show that, by using fuzzy interpolation techniques,
all market information is able to be easily integrated such that the pricing design can be transformed into an LMI problem, which is efficiently solvable due to its convexity.
Unlike the ACE pricing scheme which only employs  feedback of the imbalanced energy, the proposed pricing scheme results in better performance due to  its full utilization of   all system states, i.e., power supply, power demand, and imbalanced power.

As illustrated by simulations, the proposed design
outperforms the existing ACE pricing scheme in the following two ways: it manages the imbalanced energy more quickly;
and it is more robust against system disturbances, i.e., the uncertainty in marginal benefit and cost, and the fluctuating power input from the RES.
Despite the differences, the price in the proposed scheme
still tends to the same equilibrium as the ACE pricing scheme.
The proposed pricing scheme maintains its performance on the imbalanced energy by vibrating the price. Therefore, price vibration is crucial to balancing the power demand and power supply.

Referring to Remark \ref{rmk_market_power_sys}, some existing studies worked on interconnected systems comprising power systems and market dynamics, in which the results were established based on some specific models.
Our planned future work includes designing a practical strategy to deploy
the proposed pricing scheme in certain real-world scenarios comprising regulators and traditional utilities.
In addition, it is also of interest to consider a centralized scheme for a network of microgrids
 so that energy resources can be fully utilized from a networked system perspective, as discussed in Remark \ref{rmk_network_micro}.

\end{document}